\documentstyle[11pt,paspconf,epsf]{article}
\begin{document}
\title{A Galactic Bar to Beyond the Solar Circle and its
Relevance for Microlensing}
\author{Michael Feast}
\affil{Department of Astronomy, University of Cape Town,
Rondebosch, 7701, South Africa. \ \ email: mwf@artemisia.ast.uct.ac.za}
\author{Patricia Whitelock}
\affil{ South African Astronomical Observatory, PO Box 9,
Observatory, 7935, South Africa. \ \ email: paw@saao.ac.za}

\begin{abstract}
The Galactic kinematics of Mira variables have been studied using infrared
photometry, radial velocities, and Hipparcos parallaxes and proper motions.
For Miras in the period range 145 to 200 days (probably corresponding to
[Fe/H] in the range --0.8 to --1.3) the major axes of the stellar orbits are
concentrated in the first quadrant of Galactic longitude. This is
interpreted as a continuation of the bar-like structure of the Galactic
Bulge out to the solar circle and beyond.
\end{abstract}
\keywords {Galaxy: structure, kinematics and dynamics: Mira variables}

\section{Introduction}
  A somewhat unexpected discovery of microlensing experiments towards the
Galactic Bulge was the need to take into account a bar-like structure there
(see, e.g.\ Paczy\'{n}ski 1996). The precise nature of this bar - its
composition, extent and evolution - is only sketchily known. This is an area
where progress is likely to come from the combination of microlensing
results with other types of observations. In the present paper we summarize
the results of an analysis of the Galactic kinematics of Mira variables
based on infrared photometry, radial velocities, and Hipparcos parallaxes
and proper motions (Whitelock, Feast \& Marang 2000, Whitelock \& Feast
2000, Feast \& Whitelock 2000b). This provides evidence for a rather
extended bar-like structure in the Galaxy.

\section{Background}
Mira variables have a number of properties which makes them
particularly interesting for Galactic structure studies
\footnote {Note that the present paper refers entirely to
oxygen-rich Miras.}.

Bolometrically and in the near infrared,  e.g.\ at $K$ (2.2$\mu$m), they are
the brightest members of the old populations in which they are found (see
for instance fig 1 of Feast \& Whitelock 1987). These variables therefore
define the tip of the AGB in these populations. Because of their brightness
they can be studied with relative ease to large distances.

Observations in the LMC show that Miras follow a well defined, narrow period
- luminosity (PL) relation at $K$ or in $ M_{\rm bol}$ (Feast et al.\ 1989).
Thus Mira distances can be estimated from infrared photometry with good
accuracy.

It has long been known that the kinematics of Galactic Miras are a
function of period (see, e.g.\ fig 12 of Feast 1963). Their asymmetric
drift gets numerically larger as one moves from the longer period
Miras to the shorter period ones with a maximum of 
$\rm \sim 100\, km\,s^{-1}$ at
a period of about 175 days. This is intermediate between the value usually
quoted for the thick disc ($\rm 30 - 40\, km\,s^{-1}$, e.g.\ Freeman 1987)
and that appropriate to halo objects ( $\rm \sim 230\, km\,s^{-1}$). This
apparently intermediate population has generally been neglected in
discussions of Galactic structure and kinematics.  Judging from Miras in
globular clusters this corresponds to a metallicity, [Fe/H], range from
about --0.8 to about --1.3. As one moves to Miras of even shorter periods
the numerical value of the asymmetric drift decreases, making the shortest
period Miras similar in kinematics to Miras of much longer period. The
relationship of this shortest period group to the other Miras was long a
puzzle.

Another important property of the Miras is shown by those in globular
clusters. Miras occur in relatively metal rich clusters. If there are more
than one Mira in any cluster, their periods are all close together. Such
clusters also contain semiregular (SR) variables. In any one cluster these
have a range in periods, and in the period - luminosity plane they lie on an
evolutionary track which terminates on the Mira PL relation (see Whitelock
1986, Feast 1989). The clusters containing Miras show that for these stars
there is a good relation between period and metallicity (see fig 1 of Feast
\& Whitelock 2000a). Since the change of mass with metallicity is presumably
small for these stars, the period - metallicity relation is primarily
indicating a relation between metallicity and stellar radius at the tip of
the AGB.

These various properties have been important in using Miras
to study the Galactic Bulge itself. The work of Lloyd Evans (1976)
and others showed that there was a wide distribution of periods
amongst Miras in the Bulge. This, then, implies a wide range in metallicities.
Indeed the metallicity distribution inferred from the Mira period
distribution (Feast \& Whitelock 2000a) is similar to that derived
by Sadler et al.\ (1996) from Bulge K-type giants. The PL 
relation allows one to estimate the distances to individual stars in the
Bulge (see, e.g.\ Glass et al.\ 1995). In this way Whitelock \& Catchpole
(1992) found that Bulge Miras at positive Galactic longitudes were nearer to us
than those at negative longitudes. This was one of the first 
pieces of evidence for a barred structure of the Bulge.

\section{Present Work}
  Our present study consists of extensive  near infrared, $JHKL$, photometry
of Galactic Mira-like variables\footnote{The sample contains a few stars
classified as semiregular (SR) variables which have Mira-like properties
(see Whitelock et al.\ 2000).}, made at SAAO Sutherland, combined with
Hipparcos astrometric and photometric data and with radial velocities from
the literature.

Combining the infrared observations with the Hipparcos photometry allows us
to see clearly the cause of the anomalous result for the asymmetric drift of
the shortest period Miras discussed above. In the $ (Hp-K)_{0}$ - $
\log P$ plane (where $Hp$ is the mean Hipparcos magnitude) there are two
near parallel sequences (see Fig 1).
A main (blue) sequence and a less populated, redder, sequence which is
largely confined to the shorter periods and dominates at the very shortest
periods. In the period range where the two sequences overlap their
kinematics are quite different. The red sequence stars have kinematics which
associates them with much longer period Miras.  It is tempting therefore to
identify some at least of the stars on this red sequence as analogous to the
SRs in metal-rich globular clusters, which were discussed above, but at
somewhat higher luminosities, and to suggest that they are evolving into
longer period Miras. If this is the case we would expect the red sequence
stars to be somewhat brighter than the blue sequence ones at a given period.
This is because evolutionary tracks (at least in metal-rich globular
clusters) lie above the Mira PL relation in the period - luminosity plane.
Indeed the Hipparcos parallaxes do indicate that at a given period the red
sequence stars are brighter than the blue sequence ones. It is also possible
that some of the red sequence stars are pulsating in a higher mode than the
blue sequence stars. The following discussion refers only to the main (blue)
sequence stars. These stars may also be recognized from their infrared
colours and these show that the Miras that have been found in globular
clusters and the LMC belong to the blue sequence.
\begin{figure}
\centering
\epsfysize=7cm 
\epsfbox{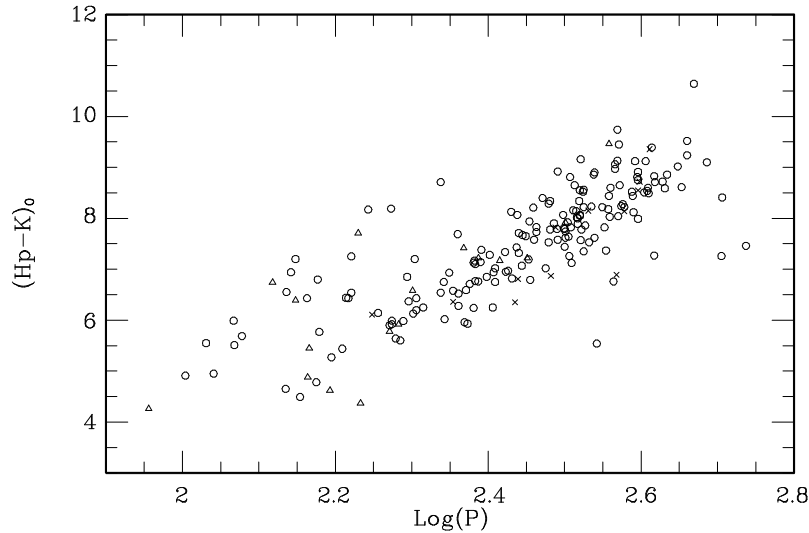} 
\caption{The period-colour relation for oxygen-rich Mira-like variables.
Stars catalogued as Miras and semi-regulars are shown as circles
and triangles, respectively. S-type stars are shown as crosses} 
\end{figure}

Hipparcos parallaxes of blue sequence stars were used to derive the 
zero-point of the PL relation. Using this calibration one can determine
a distance modulus for the LMC. This is found to be
$18.64 \pm 0.14$ mag or slightly  greater if a correction is necessary for
a metallicity difference between the local and LMC Miras. This modulus
agrees well with that determined from Cepheids and some other distance
indicators (see the summary in Feast 1999).

We have studied the Galactic kinematics of the blue sequence Miras using
Hipparcos proper motions and radial velocities from the literature. These
radial velocities include radio observations of OH and other molecules.
Distances were derived from the mean $K$ magnitudes and the Hipparcos
calibrated Mira PL relation. However, the results are rather insensitive to
the distance scale adopted.

The three components of the space motion of each star were derived in
non-rotating cylindrical co-ordinates centred at the Galactic centre.
These are $V_{R}$ radially outwards and parallel to the plane,
$V_{\theta}$ at right angles to $V_{R}$ in the direction of Galactic
rotation, and $w$ in the direction of the North Galactic Pole.
The data were analysed by dividing the Miras into six (blue
sequence) groups, according to period.

The results for $V_{\theta}$ confirm previous work. 
In the longest period group (mean period 453 days), 
$V_{\theta} = 223 \pm 4 \rm \, km\,s^{-1}$. This is only slightly
smaller than the circular velocity implied by the kinematics of Cepheids
($\rm 231\, km\,s^{-1}$, Feast \& Whitelock 1997). The value of
$V_{\theta}$ drops as the period decreases reaching $\rm 147\pm
14\,km\,s^{-1}$ in the group of mean period 175 days (17 stars). In this
calculation one Mira in the period range 145 to 200 days (S Car) has been
omitted, as it is moving on a highly eccentric retrograde orbit.

The main surprise of the analysis was in the values of $V_{R}$. This is
$\rm +67 \pm 17\,km\,s^{-1}$ in the group of mean period 175 days, whilst
for the other groups there is a small net outward velocity.

There seem to be three possible explanations of this large mean outward
radial motion in the 145 to 200 day group.

If the Galaxy is axially symmetric the result
would imply that there is a general, axially symmetric, outward, radial
motion of Miras in the period range 145 - 200 days. 
This seems unlikely.

Another alternative is that Miras in this period range belong to some
galactic interloper. Whilst this cannot be ruled out we might ask why there
are, in that case, no local stars of this type belonging to our Galaxy (such
stars exist in the Galactic Bulge itself and in Galactic globular clusters).
Also the mean motion perpendicular to the plane is small for this period
group ($w = -12\pm 12\,\rm km\,s^{-1}$) indicating that any interloper must
be moving very closely parallel to the Galactic plane.

The most likely explanation would seem to be that there is a Galactic
asymmetry leading to a deficit of incoming orbits in the
solar neighbourhood for Miras in this period range. The values of
$V_{\theta}$ and $V_{R}$ for this group then show that the major axes
of their Galactic orbits are concentrated in the
first quadrant of galactic longitude. 
All workers place the major axis of the Bulge-bar in this quadrant.
A simple first-order calculation
shows that the mean orbit has a major axis making an angle of 
$17^{+11}_{-4}$ degrees with the Sun - Centre line. This agrees with 
the position angle of the Bulge-bar derived by Binney et al.\ (1991)
from gas dynamics in the central region (16 degrees). It would be
consistent with a number of other estimates which are in the 
20 to 30 degree range, though values near 45 degrees derived
by some workers agree less well. Rough estimates suggest that the
orbits of some of these short period Miras are sufficiently eccentric to
penetrate into the Bulge region itself at perigalacticum.

\begin{figure}
\centering
\epsfysize=7cm 
\epsfbox{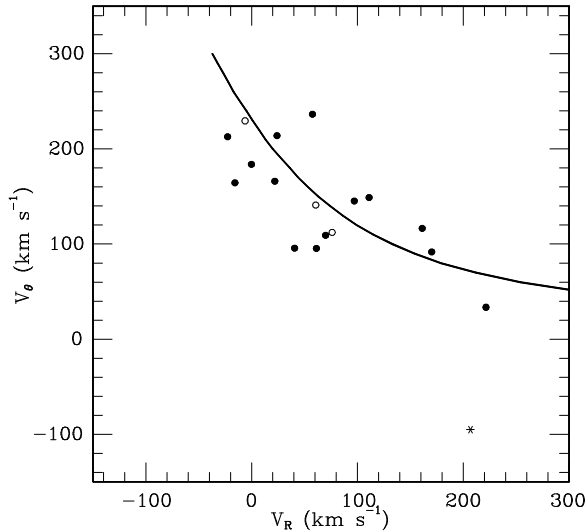} 
\caption{A plot of $V_{\theta}$ against $V_{R}$ for stars with periods
in the range 145 to 200 days. The curve is the relation expected on a simple
model in which the major axes of the orbits of all the stars are aligned at
$17^{\rm o}$ to the Sun-Centre line. Stars for which any velocity component
has a standard error in excess of $\rm 20\,km\,s^{-1}$ are represented by
open symbols. The asterisk denotes S~Car.}
\end{figure}

Figure 2 shows a plot of $V_{\theta}$ against $V_{R}$ for stars in the 145 to
200 day group. The curve is the relation that would be expected, on a simple
model, if the major axes of the orbits of all the stars were aligned at an
angle of 17 degrees to the Sun - Centre line.

The number of Miras in the short period group is small, 17 stars (omitting S
Car for reasons mentioned above).  However, the lifetime of a Mira is short
($\sim 2 \times 10^{5}$ years). Thus these stars are tracers of a much
larger population. This can be seen from the metal-rich globular clusters,
where one such cluster may contain just one Mira. It would be particularly
interesting to identify other objects associated with the 145 to 200 day
Miras. The globular cluster results suggest that such objects would have
metallicities in the range $-0.8\geq [Fe/H] \geq -1.3$ and the age of
globular clusters in this metallicity range. However, the situation may be
complex. Minniti et al.\ (1997) suggest that the RR Lyrae variables in the
Bulge do not belong to a bar-like distribution, despite the fact that the
ones in the NGC\,6522 field, at least, have $[Fe/H] \simeq -1$ (Walker \&
Terndrup 1991). It may be that choosing Miras by period allows one to sort
out a more homogeneous population than is possible in other ways.

So far as microlensing is concerned this work would seem relevant
for at least two reasons.\begin{enumerate}
\item It gives hope of defining further the composition and general
character of the Galactic bar.
\item It warns that one should not necessarily consider Galactic structure
and kinematics as ``knowns''  into which microlensing studies must
be fitted.  There may well be surprises; and microlensing data may
make an important contribution to uncovering them. \end{enumerate}

\acknowledgements This paper is based on observations made with the
Hipparcos satellite and at the
South African Astronomical Observatory (SAAO).


\begin{references}
\reference Binney, J.J., Gerhard, O.E., Stark, A.A., Bally, J.,
\& Uchida, K.I. 1991, \mnras, 252, 210 
\reference Feast, M.W. 1963, \mnras, 125, 367
\reference Feast, M.W. 1989, in: The Use of Pulsating Stars in
Fundamental Problems of Astronomy, ed. E.G. Schmidt, Cambridge
University Press, p.~205
\reference Feast, M.W. 1999, \pasp, 111, 775
\reference Feast, M.W., Glass, I.S., Whitelock, P.A. \& Catchpole, R.M.
1989, \mnras, 241, 375
\reference Feast, M.W. \& Whitelock, P.A. 1987, in: Late Stages of
Stellar Evolution, eds. S. Kwok \& S.R. Pottasch, Reidel, p.~33
\reference Feast, M.W. \& Whitelock, P.A. 2000a, in: The Chemical
Evolution of the Milky Way, eds. F. Giovannelli \& F. Matteucci,
Kluwer, in press, \\astro-ph/9911393
\reference Feast, M.W. \& Whitelock, P.A. 2000b, \mnras, submitted
\reference Freeman, K.C. 1987, ARA\&A, 25, 603
\reference Glass, I.S., Whitelock, P.A., Catchpole, R.M.,
\& Feast, M.W. 1995, \mnras, 273, 383
\reference Lloyd Evans, T. 1976, \mnras, 174, 169
\reference Minniti, D. 1997, in: Variable Stars and the Astrophysical
Returns of Microlensing Surveys, eds. R. Ferlet, J.P. Maillard 
\& B. Raban, Editions Fronti\`{e}res, p.~257 
\reference Paczy\'{n}ski, B. 1996, ARA\&A, 34, 419
\reference Sadler, E.M., Rich, R.M. \& Terndrup, D.M. 1996, \aj, 112, 171
\reference Walker, A.R. \& Terndrup, D.M. 1991, \apj, 378, 119 
\reference Whitelock, P.A. 1986, \mnras, 219, 525
\reference Whitelock, P.A. \& Catchpole, R.M. 1992, in: The Center, Bulge
 and Disk of the Milky Way, ed. L. Blitz, Kluwer, p.~103
\reference Whitelock, P.A., Feast, M.W. \& Marang, F. 2000, \mnras, submitted
\reference Whitelock, P.A. \& Feast, M.W. 2000, \mnras, submitted
\end{references}
\end{document}